# T Tauri Variability in the Context of the Beat Frequency Model

Kester W. Smith, Geraint F. Lewis & Ian A. Bonnell
*Institute of Astronomy, Madingley Road, Cambridge CB3 0HA.*

**ABSTRACT**

We examine the implications of a beat frequency modulated model of T Tauri accretion. In particular we show that measurements of the variability of accretion generated lines can be used in conjunction with existing photometry to obtain a measurement of the underlying photospheric and disc flux. This provides an independent way of checking spectral energy distribution modelling. In addition, we show how spectroscopy of T Tauri stars can reveal the inclination angle between the magnetic axis and the plane of the disc.

## INTRODUCTION

The Classical T Tauri stars (CTTS) are believed to be actively accreting material from circumstellar disks, giving rise to a pronounced veiling continuum, and characteristic emission features such as $H_\alpha$ and Ca II H + K. Periodic photometric variability is also often observed in these systems. The standard picture is that this variability is caused by rotational modulation of localised accretion shocks at the stellar surface. This model cannot be applied in all cases, however, as many periods are inconsistent with the stellar rotation as revealed by $v \sin i$ measurements (Bouvier et al 1986; Bouvier et al 1993). Furthermore, there are a number of CTTS for which the period is seen to change (Bouvier et al 1994), an impossible situation if the period reflects the stellar rotation. In a recent paper (Smith et al 1994), we have discussed the possibility that the photometric period is caused by an interaction between a misaligned magnetosphere and a clumpy disk, a model analogous to that developed for horizontal branch quasi-periodic oscillations in low-mass X-ray binaries (Alpar & Shaham 1985). The period seen in the photometric monitoring would be the beat period between the stellar rotation period and the orbital period at the inner edge of the disk, or some harmonic of this period. Support for this model has been found in line profile variability of SU Aur (Johns & Basri 1995) which displays evidence for periodic accretion with the same period as that observed in the photometry.

In this paper we point out that if the periodicity is caused by beat frequency modulation, the amplitude of the variation depends on the non-axisymmetry of the magnetosphere at the inner edge of the disc, which in turn depends solely on the inclination of the disk to the equatorial plane of the star. It is thus possible to recover the inclination of the disk plane from the photometry. In sections 2 and 3 we outline the model, in section 4 we show how the model can be applied in a specific case, and in section 5 we summarize our conclusions.

## 2 THE MODEL

The star accretes material from the inner edge of its disk as magnetic torques brake the disk material, which then falls in along the field lines. The rate of accretion will depend on the magnetic field strength, the surface density of the disk, and the exact nature of the interaction between the field and the disk.

In the beat frequency model, the interaction between the non-axisymmetric field and inhomogeneous disk leads to variations in the accretion rate, with a characteristic frequency $|\omega_{inner} - \omega_*|$, the beat frequency between the disc and the star (here $\omega_{inner}$ is the Keplerian frequency at the inner edge of the disk, and $\omega_*$ is the spin frequency of the star). Assuming the clumpy structure of the disk remains constant for a large number of orbits (a necessary assumption if the T Tauri periodicities are to be explained in this way), the amplitude of the variations in the accretion luminosity depends only on the difference between the minimum and maximum field strengths at the inner edge of the accretion disc.

The rate at which material is stripped from the disk as a function of field strength is not well known. The torque exerted on a given disk annulus by a field which fully penetrates the disk is proportional to the product of the poloidal and toroidal field components, $B_z$ and $B_\phi$ (Armitage 1995).

Here the poloidal field is assumed to be a pure dipole field, while the toroidal field is formed by distortion of the dipole field by the fast moving disk material. The magnitude of $B_\phi$ depends on which process is invoked to limit its growth. Here we follow Armitage (1995) and assume $B_\phi$ is limited by reconnection of field lines. In this case the equilibrium value of $B_\phi$ is of order $B_z$. A realistic estimate of the torque is then

$$T \propto B_\phi B_z \propto B_z^2 \tag{1}$$

In the interests of generality, we use $T \propto B_z^\gamma$ for our torque. We assume that the stripping rate and the accretion luminosity are linearly dependent on the torque. Thus, the ratio of accretion luminosity at maximum field strength, $f(B_{max})$ to that at minimum field strength, $f(B_{min})$, is

$$\frac{f(B_{max})}{f(B_{min})} = \left(\frac{B_{max}}{B_{min}}\right)^\gamma = Z \tag{2}$$

The measurement of Z can be used to gain a valuable insight into the geometry of a given T Tauri system, since it depends only on the ratio of maximum to minimum field strength at the inner edge of the disc. For a purely dipolar field, this ratio is;

$$Z^{\frac{1}{\gamma}} = \left(\frac{B_{max}}{B_{min}}\right) = (1 + 3\sin^2\theta)^{\frac{1}{2}} \tag{3}$$

(Lipunov 1987) We argued above that $\gamma = 2$ for a field whose toroidal component is limited by reconnection. Thus, knowing Z, it is possible to recover the inclination angle between magnetic dipole axis and the plane of the disc, $\theta$.

## 3 PERIODIC ACCRETION AND EMISSION LINES

By making the assumption that the dependance of accretion-generated lines on the field strength is the same as that of the continuum, we can use the ratio of maximum to minimum line flux in a number of lines to measure the ratio Z. Kuhi (1991) showed that, for a number of stars, a strong correlation exists between the flux in $H_\alpha$ and the ultraviolet continuum, which is dominated by accretion luminosity, and also between the flux in Ca II K and the ultraviolet continuum. This provides support for our assumption above. The lines used should be high energy, so they are more likely to be driven purely by the violent accretion events at the star's surface, and should vary with the same period as the continuum, which would demonstrate their link to the varying accretion.

The periodic variability in the bolometric photometry is;

$$\Delta M = -2.5 \log_{10}\left(\frac{F_0 + Zf(B_{min})}{F_0 + f(B_{min})}\right) \tag{4}$$

Here $\Delta M$ is the amplitude of the photometric variability and $F_0$ is the underlying, constant, combined luminosity from the star and disc. By making measurements of the maximum and minimum magnitude attained by the star in its periodic variations, and knowing Z from line measurements, we can solve for $F_0$ and $f(B_{min})$, and hence find $f(B_{max})$. The equations for photometry in a band, X, are

$$X_{max} = X_0 + \alpha Z f(B_{min}) \tag{5}$$

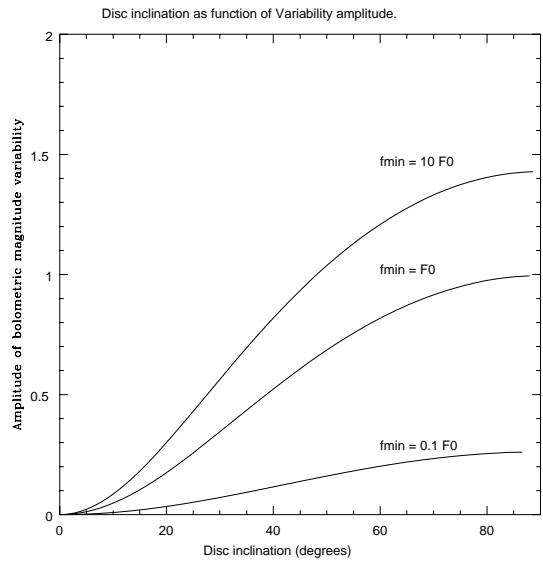

**Figure 1.** The disc-magnetic field inclination as a function of the observed bolometric variability. Three different cases are considered, with $f_{min}$ equal to the star and disc luminosity, $10F_0$, $F_0$, and $0.1F_0$, corresponding roughly to accretion rates of $10^{-6}$, $10^{-7}$ and $10^{-8}$ $M_\odot yr^{-1}$ In all cases $\gamma$ is taken to be 2.

and

$$X_{min} = X_0 + \alpha f(B_{min}) \tag{6}$$

Where $X_{max}$ and $X_{min}$ are the maximum and minimum X band absolute fluxes, $X_0$ is the X flux from the star plus the disc, and $\alpha$ is the fraction of the accretion luminosity appearing in the X band. Combining these equations gives us

$$\frac{X_{max} - X_0}{X_{min} - X_0} = Z \tag{7}$$

Knowing these fluxes from the photometry, and Z from the line variability it is possible to solve for $X_0$. It is then possible to solve for $\alpha f(B_{min})$.

Broadband photometric observations of periodic variability have been published by a large number of authors (Bouvier et al 1993; Bouvier et al 1994; Attridge & Herbst 1992). The amplitude of variability increases sharply in the B and U bands, possibly because the non-variable contribution of the stellar photosphere is much lower here. Typical observed amplitudes are 0.5 magnitudes in U, decreasing to 0.1 magnitudes in V, and less in R and I.

Figure 1 shows the amplitude of the observed bolometric variability produced by different field inclinations. The three curves represent different accretion rates, and hence different accretion luminosities. It can be seen that the amplitudes produced are consistent with the observed photometry for relatively modest inclination angles.

In addition to directly affecting the accretion and the emission lines involved, the inclined rotating magnetic field will also affect the surrounding disc and its emission.

First of all, magnetospheric accretion removes the inner part of the disc up to the magnetospheric radius. With an inclined dipole, the disc sees a varying magnetic field strength. Crudely translated into a varying magnetospheric radius, a dipolar field inclined at 30 degrees implies a variation in the magnetospheric radius of 17%. Although this in practice won't translate directly into a change of the inner hole of the disc, as the disc will not adapt immediately to the change in the magnetic field, it will have effects on the emission from the inner part of the disc. Furthermore, the dissipation of the magnetic field in the disc will increase the emission in the near infra-red (Armitage & Clarke 1995; Kenyon, Yi & Hartmann 1995). Any asymmetries in the disc will then also produce variability in the near infra-red with the same period as the accretion. This might explain the near infra-red variability found in T Tauri stars on timescales from a few days to weeks (Hillenbrand, personal communication).

| Line | λ (A) | EW min | EW max | Z |
|---|---|---|---|---|
| Ca II H | 3968 | -27 | -60 | 2.22 |
| Ca II | 8662 | -50 | -68 | 1.36 |
| $H_\alpha$ | 6561 | -75 | -105 | 1.4 |
| $H_\beta$ | 4861 | -24 | -42 | 1.75 |
| He I | 5876 | -2 | -3.8 | 1.9 |

**Table 1.** Equivalent width variations for five lines in the spectrum of DG Tau, from Guenther (1994). The implied vales of Z are given, assuming a one to one correlation between the equivalent width and line flux.

| Band | Amp | $\frac{F_{max}}{F_{min}}$ | $F_0$(%) | DG Tau(M0) | DG Tau(K7) |
|---|---|---|---|---|---|
| U | 0.5 | 1.58 | 42 | 1 | 20 |
| B | 0.4 | 1.45 | 55 | 23 | 70 |
| V | 0.35 | 1.38 | 62 | 33 | 91 |
| R | 0.3 | 1.32 | 68 | 79 | 100 |
| I | 0.22 | 1.22 | 78 | - | - |

**Table 2.** Calculation of underlying photospheric and disc fluxes in each band for the T Tauri star DG Tau, and comparison with values estimated from an SED published by Basri & Bertout (1989) and photospheric templates for M0 and K7 dwarfs. Column two shows the amplitude of the variability in each band, column three the corresponding ratio of maximum to minimum flux. In column four we show the fraction of the flux attributable to underlying photosphere plus disk flux in Basri & Bertout's SED. Columns five and six show the values obtained using our calculations, column five assuming an M0 photosphere, and column six a K7 photosphere.

## 4 AN ILLUSTRATIVE EXAMPLE

Two sets of observations are necessary to complete the measurements we propose above. Firstly, a spectroscopic monitoring of high excitation energy accretion linked emission lines to determine Z, and secondly, a measurement of the photometric variability.

Any emission lines observed would need to be linked to the accretion process. For this reason they should be seen to vary with the same period as the photometry. Such variability was observed by Guenther (1994), for 5 emission lines in three systems. Selecting high excitation energy lines is necessary, since such lines are expected to form in the hot accretion shocks, and are thus more likely to be purely accretion driven.

DG Tau is a CTTS believed to have an M0 photospheric spectrum. It was one of the three stars studied by Guenther (1994). We can use this spectroscopy, together with photometry which exists for this star, to work through the steps we outline above and thus determine the appropriate parameters. It should be noted that the spectroscopy and photometry are not simultaneous, and further observations would be necessary to fully vindicate the model. Guenther's spectra were not flux calibrated, so the flux variability needed to calculate Z is not directly available. Line fluxes should ideally be used for this measurement, however, Guenther notes that the variations in the equivalent widths of these lines for DG Tau are mostly due to variations in the line strengths, rather than continuum variations. In view of these considerations, we feel justified in using equivalent widths here to calculate Z. The approximate peak to peak variations of the lines studied by Guenther are shown in Table 1

We will neglect Ca II 8662 and $H_\alpha$ as they are lower energy, and less likely to be purely accretion driven. Taking a simple average of the other lines gives us $Z = 1.96$. From equation 3 assuming $\gamma = 2$, we can deduce that this implies an angle of 34 degrees between the axis of the dipole field and the plane of the disc. Photometry of DG Tauri taken by Bouvier et al (1993) shows periodicity in U,B,V,R and I bands. Applying the equation above to the amplitudes of the periodic variations found by them and using $Z = 1.96$ gives the results shown in Table 2 Also shown are the values obtained by comparing a spectrum of DG Tau from a paper by Basri & Bertout (1989) to an M0 photosphere, as used by them, and to a K7 photosphere, also taken from their paper.

It was not possible to obtain figures for the I band, as the published data do not extend far enough into the red. It can be seen from Table 2 that our values lie between those for the two photospheres. We therefore conclude that our method predicts that the photosphere of DG Tau is somewhere between an M0 and a K7, possibly a K8.

## 5 CONCLUSIONS

We have discussed some consequences of the recently proposed beat frequency modulation model of T Tauri accretion, and showed how these consequences can be used to determine certain parameters of individual systems.

The method outlined above provides a measurement of the underlying stellar and disc fluxes, independent of any spectral energy distribution (SED) modelling. This in itself is a very useful measurement, as it allows more tests to be applied to such SED calculations, and will also help to constrain the positions of T Tauri stars on the HR diagram. This is perhaps most important in determining T Tauri ages.

The model in this paper presents the first measurement of the relative orientation of the axis of the magnetic field and the disk in a T Tauri system. As illustrated in Section 3, this orientation may have a profound effect on the inner structure of the accretion disk and infrared variability. These orientation diven effects, and their observables are currently a topic of further research and will be presented at a later date.

Magnetically controlled accretion in T Tauri Stars has recently received much supporting evidence, both observationally and as a model to explain their slow rotation rates (Edwards 1995). It has also been hypothesized that the stellar magnetic field may be relevant in driving a stellar wind or jet (Edwards 1995; Shu et al 1994). In all of these models, the orientation of the magnetic field is assumed, for simplicity, to be perpendicular to the surrounding disc. The positive identification of a T Tauri Star undergoing periodic accretion at the beat frequency of the star and disc will yield the orientation of the magnetic field relative to the surrounding disc. This will be important not only as a constraint on the models of magnetospheric accretion and their possible relation to stellar winds but also as a constraint on the physics of magnetic fields in pre-main sequence stars.

The search for periodicities in emission line strengths is still in its early stages. These lines are strongly linked with other accretion signatures, however, and so it is to be expected that they will have such periodicities in common with the continuum. Recent studies appear to bear out thois expectation (Guenther 1994) A spectroscopic campaign is currently being planned which will as part of its aim, attempt a measurement of Z for a small number of T Tauri stars.

## 6 ACKNOWLEDGEMENTS

We thank Eike Guenther, Lynne Hillenbrand, Philip Armitage, Melvyn Davies and Cathie Clarke for helpful discussions. We also thank the computer support staff at the Institute of Astronomy for maintaining the machines used in this work. KWS acknowledges a PPARC studentship.

## REFERENCES

Alpar M.A. & Shaham J., 1985, Nature, 316, 239.
Armitage, P. 1995, MNRAS, 274, 1242.
Armitage, P. & Clarke C.J., 1995, MNRAS, In press.
Attridge J. & Herbst W., 1992, ApJLett 398, L61.
Basri G. & Bertout C., 1989, ApJ, 341, 340.
Bouvier J., Bertout C., Benz W.,, Mayor M. 1986, A&A, 165, 110.
Bouvier J., Cabrit S., Fenandez M., Martin E., Matthews J., 1993, A&A, 272, 176.
Bouvier J., Covino E., Kovo O., Martin E.L. ,Matthews J.M., Terranegra L., Beck S.C., 1994, A&A, In press.
Bertout C., Basri G., & Bouvier J., 1988, Ap.J. 330 350.
Edwards S., 1995, Revista Mexicana de Astronomia y Astrofisica, 1995, In press.
Guenther E., 1994, in Disks and Outflows around young Stars, ed. Beckwith S., In press.
Johns C.M. & Basri G., 1995, Ap.J., In press.
Kenyon S.J., Yi I. & Hartmann L., 1995, Ap.J. In press.
Kuhi L.V., 1974, A&A S. 15,47.
V.M.Lipunov, 1987, Astrophysics of Neutron Stars, Springer-Verlag, pp 82-86.
Shu F., Najita J., Ostriker E., Wilkin F., Ruden S., Lizano S., 1994, Ap.J. 429, 781.
Smith K.W., Bonnell I.A. & Lewis G.F. 1995 MNRAS in press.